\documentstyle[12pt]{article}
\begin{document}
\begin{center}
{\Large {\bf Radiation in Yang-Mills formulation of gravity\\
and a generalized pp-wave metric }}\\[1cm]
{\normalsize S. Ba{\c s}kal}\\
{\small Department of Physics, Middle East Technical
University, Ankara 06531, Turkey}\\[15mm]
\end{center}
\begin{abstract}
The variational methods implemented on a quadratic Yang-Mills type
Lagrangian yield two sets of equations interpreted as the field
equations and the energy-momentum tensor for the gravitational field.
A covariant condition is imposed on the energy-momentum tensor to
represent the radiation field.  A generalized pp-wave metric is found
to simultaneously satisfy both the field equations and the radiation
condition.  The result is compared with that of Lichn\'{e}rowicz.
\end{abstract}
Based on its gauge structure which was first recognized by Utiyama
\cite{uti56}, alternative formulations of gravitation especially the
ones that are derivable from quadratic Lagrangians of Yang-Mills (YM)
type have received substantial interest in the literature~
\cite{fair77,der81,cam75}, not merely as an academic curiosity, but
essentially because when the gravitational Lagrangian is coupled to matter
the renormalization problems are much less severe \cite{stel77}.

For massless particles like photons and neutrinos radiation
presents itself, if for all observers their respective energy
flows in the same direction with that of light.  As to the
electromagnetic field this situation is well known and the neutrino
radiation in curved space is also well established \cite{aud77}.
It has been shown that any solution to Einstein-Weyl equations represent
neutrino radiation if the energy-momentum tensor of the Weyl field
can be expressed in terms of a null four vector which is collinear to
its four-momentum.  Therefore, there is no reason not to expect that
the gauge quanta of gravitation retains the same conditions to be
in a radiative state.

In this article, we begin with a YM type Lagrangian which is
quadratic in the Riemann tensor and employ Palatini's variational method
to derive two sets of equations whose interpretations are given as the
field equations and the gravitational energy-momentum tensor.  
According to Palatini's method, which was first implemented by
Stephenson\cite{step58} and then elaborated by Fairchild\cite{fair77},
the equations may be derived by independent variations of the Lagrangian
for the connection and for the metric.  We introduce covariant conditions
on the energy-momentum tensor to represent gravitational radiation
and for the solutions specialize on a generalized form of a metric
which represents plane fronted waves with parallel rays (pp-wave).
We compare the radiation criteria with that of Lichn\'{e}rowicz.

The dynamical equations to be considered here are determined by a
variational principle from the gauge-invariant action $I=\int_{M}\,L$,
where $M$ is a four dimensional spacetime endowed with a metric of $+2$
signature, and the Lagrangian $L$ is:
\begin{equation}
L=\sqrt{-g}\,R^{\mu}\,_{\nu\,\rho\sigma}\,R^{\nu}\,_{\mu}\,^{\rho\sigma}.
\label{Lag}
\end{equation}
Variation with respect to the connection
$\delta L / \delta \Gamma^{\nu}\,_{\alpha\beta}$ gives
\begin{equation}
\nabla_{\mu} R^{\mu}\,_{\nu\alpha\beta}=0,
\label{fe1}
\end{equation}
and the Bianchi identity
\begin{equation}
\nabla_{[\mu} R_{\nu\sigma]\,\alpha\beta}=0
\end{equation}
follows from the definition of the Riemann tensor.
Variation of the action with respect to the metric is defined
and interpreted as the energy-momentum tensor of the
corresponding field by many authors \cite{cam75,fair77,bas93}:
\begin{equation}
\delta g^{\mu\nu}(\frac{\delta L}{\delta g^{\mu\nu}})
\equiv \delta g^{\mu\nu}T_{\mu\nu}.
\end{equation}
Here the tensor $T_{\mu\nu}$ is symmetric and takes the form
\begin{equation}
T_{\mu\nu} = R_{\mu\kappa}\,^{\rho}\,_{\sigma}
R_{\nu}\,^{\kappa\sigma}\,_{\rho}
-\frac{1}{4}g_{\mu\nu}R^{\kappa}\,_{\tau\rho\sigma}
R^{\tau}\,_{\kappa}\,^{\rho\sigma}.
\label{en1}
\end{equation}

We define the radiation field as any solution of (\ref{fe1}), whose
energy (\ref{en1}) flows for all observers pointwise in the same
direction with the velocity of light.  That is
$T_{\mu\nu}U^{\nu} \sim l^{\mu}$, where $l_{\mu}l^{\mu}=0$ for all
$U^{\mu}$ with $U^{\mu}U_{\mu}=1$ and $\sim$ stands for proportionality.
Therefore the radiation field is represented through a condition on the
energy-momentum tensor as
\begin{equation}
T_{\mu\nu}=\rho(x)l_{\mu}l_{\nu}
\label{rc}
\end{equation}
where
$\rho(x)>0$ can be elucidated as the energy density.  It immediately
follows that the dominant energy conditions:
$T_{\mu\nu}U^{\mu}U^{\nu} \geq 0$ and $T_{\mu\nu}U^{\mu}$ to be
non-spacelike are satisfied.  From (\ref{en1}) it is seen that
$ T_{\mu\nu} $ is traceless and therefore, the condition on
$ l_{\mu} $ to be an isotropic vector is already inherent in the
description.  The trajectories of the vector field $ l_{\mu} $ are
interpreted as gravitational rays.

The criterion for the existence of gravitational radiation
proposed by Lichn\'{e}rowicz is based on an analogy with
electromagnetic radiation and imposes algebraic
conditions on the Riemann tensor as \cite{lich62}:
\begin{eqnarray}
l_{[\mu}R_{\nu\sigma]}\,_{\alpha\beta}=0,\qquad
l^{\mu}R_{\mu\nu\,\alpha\beta}=0,          \label{lich}
\end{eqnarray}
with $l_{\mu}\neq 0$ and $ R_{\mu\nu\,\alpha\beta}\neq 0 $.  However,
contracting the firs expression with $ R^{\nu\sigma\kappa\tau} $ and making
use of the second one, it is seen that these conditions yield to the
vanishing of $T_{\mu\nu}$ and therefore the concept of energy transfer
becomes ambiguous.

To gain insight into the properties of the radiation field
we consider a more general form a pp-wave metric:
\begin{equation}
ds^{2}=2\,du\,dv+dx^{2}+dy^{2}+2\,h(v,x,y,u)\,du^{2} \label{met}.
\end{equation}
Unlike the function in the metric that represents plane-fronted waves
with parallel rays waves whose dependence is only on $x, y$ and
$u$~\cite{kundt62}, here the metric fuction $h$ depends on all of
the coordinates.  The null vector field $ l^{\mu}=\delta^{\mu}\,_{1} $
is subject to
\begin{equation}
\nabla_{\mu}\,l^{\nu}=\alpha_{\mu}\,l^{\nu}   \label{recur}
\end{equation}
for some vector $ \alpha_{\mu} $.  Contracting (\ref{recur})
with $ l^{\mu} $ yields
\begin{equation}
l^{\mu}\nabla_{\mu}\,l^{\nu}=\kappa\,l^{\nu}
\end{equation}
expressing that the trajectories of the vector field $ l^{\mu} $ are
null geodesics.  This vector field can be chosen as
$ l_{\mu}=\partial_{\mu} S $ for some scalar $ S $.  It then follows
that all of the optical parameters determined by the geodesic null
congruence vanish.  The function $S$ is a solution of the eikonal equation
$ g^{\mu\nu}\,\partial_{\mu}S\,\partial_{\nu}S=0 $
and the hypersurfaces $S=constant$ represent the gravitational wave
fronts, and are identical with the characteristics of Einstein's vacuum
and Maxwell's equations \cite{li3}.
Moreover, the function $ S $ obeys
$g^{\mu\nu}\,\nabla_{\mu}\nabla_{\nu}S=0,$
which is the well-known wave equation.

The field equation (\ref{fe1}) evaluated over (\ref{met}) reduce to:
\begin{eqnarray}
h_{vvv}=0,\qquad h_{vvx}=0, \qquad h_{vvy}=0        \\
h_{vxy}=0, \qquad  h_{vxx}=0, \qquad    h_{vyy}=0   \label{fs} \\
\partial_{v}\,(h_{xx}+h_{yy}+h_{vu})=0       \\
\partial_{x}\,(h_{xx}+h_{yy}+h_{vu}) + h_{v}\,h_{vx}-h_{x}\,h_{vv}=0
\label{ei} \\
\partial_{y}\,(h_{xx}+h_{yy}+h_{vu}) + h_{v}\,h_{vy}-h_{y}\,h_{vv}=0,
\label{ni}
\end{eqnarray}
and the surviving components of the energy-momentum tensor are:
\begin{eqnarray}
\begin{array}{llll}
T_{12}=-h_{vv}^{2},   &  T_{22}=h_{vv}^{2},
&   T_{23}=-2\,h_{vv}\,h_{vx},   &    T_{33}=h_{vv}^{2}, \\
& T_{34}=-2\,h_{vv}\,h_{vy}, &
 T_{44}=2\,(h_{vx}^{2}+h_{vy}^{2}-h\,h_{vv}^{2}). & \\
\end{array}
\label{cetg}
\end{eqnarray}
A closer examination reveals the tensorial relations
\begin{equation}
T_{\mu\nu}=-R\,P_{\mu\nu}+\rho\,l_{\mu}\,l_{\nu}  \label{up}
\end{equation}
and
\begin{equation}
T_{\mu\nu}=\lambda\,g_{\mu\nu}+l_{\mu}k_{\nu}+l_{\nu}k_{\mu},
\end{equation}
where the scalars $ \lambda,\,\rho $ and the vector $ k_{\mu} $ can
easily be determined.  Here $R$ is the curvature scalar and $P_{\mu\nu}$
is the traceless Einstein tensor.
For comparative reasons we calculate the Einstein tensor
$S_{\mu\nu}=R_{\mu\nu}-\frac{1}{2}g_{\mu\nu}R$, which can be
expressed as
\begin{equation}
S_{\mu\nu}=-\frac{1}{R}\,T_{\mu\nu}-\frac{1}{4}\,R\,g_{\mu\nu}
+\frac{\rho}{R}\,l_{\mu}l_{\nu}.
\end{equation}
and observe that for this particular metric the right hand
side is quite different from all known forms of energy appearing
in Einstein's equations.

For the metric in (\ref{met}) to represent a radiation field,
we first let $ h_{vv}=0 $ with $ \rho\,>\,0 $, where
\begin{equation}
\rho=2(h_{vx}^{2}+h_{vy}^{2}). \label{ro}
\end{equation}
Therefore, $ h $ is of the form $h=B(x,y,u)\,v+C(x,y,u),$ where
$ B $ and $ C $ are independent of $ v $, and $ B $ is not a
function of $ u $ only.  Then, as $ l_{\mu}=\delta ^{4}_{\mu},$ the
gravitational energy tensor in (\ref{en1}) now takes the appropriate
form (\ref{rc}).  Field equations impose more conditions on the
functions $B$ and $C$.  From (\ref{fs}) it is seen that
$B=\alpha(u)x+\beta(u)y+\gamma(u)$ and so (\ref{ei}) and (\ref{ni})
reduce to: $F_{x}=-B\,B_{x}$ and $F_{y}=-B\,B_{y}$, where
$F=C_{xx}+C_{yy}+B_{u}$. These two can be written as:
\begin{equation}
\alpha\,F_{x}-\beta\,F_{y}=0.
\end{equation}
The characteristic system associated with this equation is:
\begin{equation}
\frac{dy}{\alpha}=\frac{dx}{-\beta}=\frac{du}{0}=\frac{df}{0}.
\end{equation}
It admits the first integrals $u,\,\,\alpha\,x+\beta\,y+\gamma,\,\,F$,
and hence $F$ is of the form $F=F(B,u)$ and $C$ is now any solution
of $C_{xx}+C_{yy}=F-B_{u}$.  Therefore, we have found the form of the
metric which simultaneously satisfies the field equations and the
radiation condition on $T_{\mu\nu}$.  With this solution, the scalar
$ \rho $ in (\ref{ro}) remains constant along the geodesic null
congruence and this leads to the conservation of the gravitational
energy tensor.

It is worthwhile to study the algebraic properties of the conformal Weyl
tensor with respect to the principal null vector $ l_{\mu} $.
With $ l^{\mu}=\delta^{\mu}\,_{1} $ the following
characterization for different Petrov types are obtained:
\begin{equation}
\begin{array}{llll}
R=0, & h_{vx}^{2}+h_{vy}^{2}=0, & h_{xy}^{2}+(h_{xx}-h_{yy})^{2}=0
                                & \Leftrightarrow\;\;\mbox{type O},\\
R=0, & h_{vx}^{2}+h_{vy}^{2}=0, & h_{xy}^{2}+(h_{xx}-h_{yy})^{2}\neq 0
                                & \Leftrightarrow\;\;\mbox{type N},\\
R=0, & h_{vx}^{2}+h_{vy}^{2}\neq 0, &
                                & \Leftrightarrow\;\;\mbox{type III},\\
R\neq0, & & & \Leftrightarrow\;\;\mbox{type II or D}.
\end{array}
\end{equation}

We have presented a covariant formulation of gravitational radiation based
on the energy tensor of the gavitational field derived from a YM
type Lagrangian.  As a specific example for the illustration of a radiation
field, a space-time metric admitting plane waves is considered.
The plane wave is strictly parallel (i.e., $\alpha$ in (\ref{recur}) is zero)
if and only if $ h $ is independent of $ v $.  Then this metric becomes
identical with the one corresponding to pp-waves, which represents radiation
in the sense of Lichn\'{e}rowicz.  However, we note that in this case,
since $T_{\mu\nu}=0$, the concept of energy transfer becomes ambiguous.
The algebraic properties of the Weyl tensor are studied.  We also show
that this metric is never algebraically general, and that radiation
corresponds to type III.

I would like to thank to F. {\"O}ktem for suggesting the radiation
problem and for valuable discussions, and to T. Dereli for bringing
many of the relevant papers into my attention.

\end{document}